\documentclass[nofootinbib,superscriptaddress,onecolumn,preprintnumbers]{revtex4}
\usepackage{graphicx}
\usepackage{amsmath,amssymb}
\usepackage[normalem]{ulem}
\renewcommand{\eqref}[1]{Eq.(\ref{#1})}

\usepackage{color}

\begin{document}
\title{Running Hubble Constant with the Redshift\\ as a Marker of Evolutionary Dark Energy}

\author{N. Carlevaro}
\affiliation{Nuclear Department, ENEA - C. R. Frascati, Via E. Fermi 45, 00044 Frascati, Italy}

\author{G. Montani}
\affiliation{Nuclear Department, ENEA - C. R. Frascati, Via E. Fermi 45, 00044 Frascati, Italy}
\affiliation{Physics Department, ``Sapienza'' University of Rome,  P.le Aldo Moro 5, 00185 Roma, Italy}

\author{M.G. Dainotti}
\affiliation{Division of Science, National Astronomical Observatory of Japan, 2-21-1 Osawa, Mitaka, Tokyo 181-8588, Japan}
\affiliation{The Graduate University for Advanced Studies (SOKENDAI), Shonankokusaimura, Hayama, Miura District, Kanagawa 240-0115}
\affiliation{Space Science Institute, Boulder, CO, USA}

\begin{abstract}
We discuss the interpretation of an observed running Hubble constant with the redshift, in terms of the possible underlying physical scenario. Assuming, like in the $\Lambda$CDM model, that the matter and dark energy components do not directly interact, we arrive at the conclusion that the physical content of the cosmological dynamics can always be represented with an evolutionary dark energy paradigm. We discuss both the case of the power-law running Hubble constant, and that of an optimized fitting model, both studied with the binned Pantheon Sample of the Type Ia Supernovae data. These two cases turn out to be associated with a phantom-like evolutionary dark energy. 
\end{abstract}

\maketitle

The last decade of cosmological studies \cite{2021CQGra..38o3001D} and observations \cite{1998AJ....116.1009R,1999ApJ...517..565P,Planck:2018vyg,Scolnic_2018,Scolnic_2022,Brout:2022vxf,Dainotti2021apj-powerlaw,arx2206.11447,DESI:2025,DESI:2025zgx} demonstrated the emergence of a new era, characterized by tensions in the data, 
among which the so-called ``Hubble tension'' \cite{2021CQGra..38o3001D,riess18,2021ApJ...908L...6R} stands out (i.e. more than $5\sigma$ discrepancy between the Hubble constant value detected by the Planck and the SH0ES Collaborations), as well as evidence for an evolutionary dark energy 
outlined by the DESI Collaboration (see also \cite{dainotti-desi}). 

These new issues of modern cosmology call attention to both  
possible astrophysical calibration effects and new physics underlying the cosmological 
dynamics, see \cite{2021ApJ...914L..40D} and \cite{2020PhRvD.102b3518V,schiavone_mnras,Montani:2023xpd,Montani_2025}, respectively. Independently of the real origin of the observed phenomenology, various authors investigated 
the possibility of an effective running Hubble constant as a marker to outline deviation from the $\Lambda$CDM model or, more in general, from a given reference model \cite{Dainotti2021apj-powerlaw,Dainottigalaxies10010024,2025JHEAp..4800405D,fazzari2025,arx2203.10558,arx2206.11447,Colgain2021DDE,Schiavone:2022shz,2024MNRAS.530.5091X,Liu2025PhRvD112L3539}.  When the running Hubble constant is postulated from a phenomenological point of view or deduced from the data analysis, its 
physical meaning does not remain 
clearly determined. However, 
for the power-law behavior 
obtained as a good data fit 
of the Pantheon \cite{Dainotti2021apj-powerlaw,Dainottigalaxies10010024} and the so-called Master \cite{2025JHEAp..4800405D} samples 
of the Type Ia Supernovae (SNeIa), a possible physical explanation has 
been proposed in \cite{schiavone_mnras,montaniEntropy}.

In this Letter, we face the general question of physically 
interpreting the possibly detected effective running Hubble constant. We consider this situation as a restated Friedmann equation applying the 
validity of the Bianchi identities. As a result, we arrive at the key claim that the effective running Hubble constant, 
which can emerge from observations, always marks the presence 
of an evolutionary dark energy 
fixing its equation of state. Furthermore, 
we consider an optimized model for the running function form. For these cases and the power-law model, we 
find that the nature of the 
late Universe dark energy is 
phantom-like (for early Universe dark energy scenarios see 
\cite{2023PDU....4201348P,2019PhRvL.122v1301P,2023ARNPS..73..153K,2023PhRvD.108d3513H} and for mixed early and late phenomenology see 
\cite{2023Univ....9..393V,DiValentino:2019ffd,DiValentinodark,Alestas2020PhRvD.101l3516A}). 

It is important to stress that our result relies on the idea/assumption that the dark energy and matter components do not directly interact. Despite this is 
the situation of the $\Lambda$CDM base model and it is also a 
natural assumption, scenarios for the interaction of these two effects have been 
studied in the literature \cite{Wang_2016,Montani_2025,MONTANI2024101486,2025PDU....4901965D,Yang:2018uae} (for the case in which dark energy is described via metric $f(R)$ gravity, see 
\cite{2025arXiv251220193M,Montani:2023xpd,NOJIRI201159,2007PhRvD..76f4004H}).

If the Hubble constant proves sensitive to the redshift range of the adopted sources for its determination \cite{arx2206.11447,Dainotti2021apj-powerlaw,Dainottigalaxies10010024,2025JHEAp..4800405D,fazzari2025,2026JHEAp..5300612V,Schiavone:2022shz,2010JCAP...02..008S,2018JCAP...04..051G,2018PhRvD..97l3507W}, this leads to the introduction of the effective running Hubble constant $H_0\to\mathcal{H}_0(z)$, when a base $\Lambda$CDM model is taken into account (here, $z$ is the 
ordinary redshift value and $H(z)$ the Hubble parameter with $H_0\equiv H(z=0)$). In what follows, we consider a flat isotropic Universe and we use as time variable the quantity 
$x\equiv \ln (1+z)$. The Universe expansion rate is defined as $E(x)\equiv H(x)/H_0$ and, eventually, we use the matter 
$\Omega_m$ (sum of the baryonic and dark components) and dark energy $\Omega_{DE}$ dimensionless density parameters, respectively, with the corresponding equation of state parameters $w_m$ and $w_{DE}$. The effective running Hubble constant 
is defined as $\mathcal{H}_0(x) \equiv H_0 h(x)$, where $h(0)=1$ by construction. In this scheme, a fiducial model (denoted with barred quantities) $\bar{E}^2(x)= \bar{\Omega}_m(x)+\bar{\Omega}_{DE}(x)$, is replaced by 
\begin{align}
E^2(x) = h^2(x)\big[\bar{\Omega}_m(x)+\bar{\Omega}_{DE}(x)\big]
\;,
\label{nii111}
\end{align}
which corresponds to a revised Friedmann equation.
Following the works introduced above, we consider this fiducial scheme as a base $\Lambda$CDM model, thus implying that
\begin{align}
\bar{\Omega}_m(x)=\bar{\Omega}_{m}^{0}e^{3x}\;,
\qquad
\bar{\Omega}_{DE}(x)=1-\bar{\Omega}_{m}^{0}\;.   
\label{nii356}
\end{align}

We want now to translate such a dynamics of the running Hubble constant parametrization into a physical evolutionary dark energy scheme. We thus consider that Eq.(\ref{nii111}) is rewritten as 
\begin{align}
E^2(x) = \Omega_m(x) + \Omega_{DE}(x) \;.
\label{nii2222}
\end{align}
with the natural $ii$-component of the Einstein equations that reads
\begin{align} 
(E^2)' - 3 E^2 =  3\big[ w_{m}(x) \Omega_{m}(x)+ w_{DE}(x) \Omega_{DE}(x)\big]\;,
\label{nii3244}
\end{align}
where the prime denotes differentiation with respect to $x$. In this mapping, we consider independently evolving physical density parameters, as expected for non‑interacting components. The matter density parameter $\Omega_m(x)$ thus evolves according to its own conservation law. In the following, we treat the case of pressureless matter imposing $w_m=0$, thus $\Omega_m' = 3(1+ w_m(x))\Omega_m$ provides
\begin{align}
\Omega_m(x)&=\Omega_{m}^{0}e^{3x}\;,\\   
\Omega_{DE}(x)&=E^2-\Omega_{m}^{0}e^{3x}\;.
\label{nii3444112}
\end{align}

Using these expressions, in order to make clear the dynamical implications of this approach, we substitute the $h$ dependent parametrization of $E(x)$ in Eq.(\ref{nii111}), into Eq.(\ref{nii3244}) (implementing the conditions $w_m=0$ and $\bar{\Omega}_{DE}'=0$), thus getting
\begin{align}
w_{DE}(x) =
\frac{2(\bar{\Omega}_{m}^{0}e^{3x}+1-\bar{\Omega}_{m}^{0})\,h\,h' - 3(1-\bar{\Omega}_{m}^{0})\,h^2}{3\Big[(\bar{\Omega}_{m}^{0}e^{3x}+1-\bar{\Omega}_{m}^{0})\,h^2 - \Omega_{m}^{0}\,e^{3x}\Big]}\;,
\label{nii37}
\end{align}
which describes the dynamical features of the dark energy equation of state parameter $w_{DE}(x)$ in terms of the running Hubble constant function $h(x)$ for a base $\Lambda$CDM scheme. We also recall that, from Eq.(\ref{nii3444112}), the evolutionary dark energy expression rewrites as
\begin{align}
\Omega_{DE}(x) = h^2(x)(\bar{\Omega}_m(x)+1-\bar{\Omega}_{m}^{0}) - \Omega_{m}^{0}e^{3x}\;.
\label{nii51}
\end{align}

It is immediate to notice that, by construction, the evolution of $E^2(x)$ does not depend on $\Omega_{m}^{0}$. This standard dark degeneracy (typically observed when reconstructing $w_{DE}(x)$ in a non-parametric way) can be fixed by analyzing the $\Lambda$CDM limit: since the whole scheme must be valid for any functional form of $h(x)$, when addressing the limiting case $h(x)=1$, we are stating that the physical model of the Universe reduces to a $\Lambda$CDM one, which of course implies $w_{DE}=-1$. From Eq.(\ref{nii37}), we obtain
\begin{align}
w_{DE}(x)|_{h(x)=1}=-1 \;\;\;\Longrightarrow\;\;\;
\Omega_{m}^{0} = \bar{\Omega}_{m}^{0}\;.
\label{nii9876}
\end{align}
Only under this condition, the ``no running'' case genuinely reduces to the cosmological constant. This formally fixes the degeneracy and closes the mapping between a running Hubble constant and the evolutionary dark energy dynamics.


In what follows, we exploit the analysis strategy of \cite{Dainotti2021apj-powerlaw,Dainottigalaxies10010024} constructing binned data of the Pantheon compilation \cite{Scolnic_2018}, which includes 1048 SNeIa collected from multiple observational surveys. In fact, such a representation shows a gradual decline of $H_0$ with increasing redshift and we test the mapping approach described above by fitting the running function $h(x)$. The dataset collection is split into 40 equi-populated redshift bins ($\sim$26 SNeIa each), balancing statistical robustness and resolution. Each bin is represented by its mean redshift, and the fit uses the binned $H_0$ values with their errors; the equi-populated scheme shifts bin centres slightly toward lower $z$. For alternative binnings in $\log z$ and $\log_{10}(1/(1+z))$ see \cite{2025JHEAp..4800405D,Dainotti2024PDU....4401428D,Dainotti2024Galax..12....4D}, and for non-Gaussian likelihoods \citep{DAINOTTI202430}. The observed distance modulus is $\mu_{\text{obs}} = m_B - M$, averaged following \cite{2010A&A...523A...7G} and \cite{2011A&A...529L...4C}, while $\mu_{\text{th}} = 5 \log_{10} d_L(z, H_0, \dots) + 25$ assumes a fiducial $\Lambda$CDM model with peculiar-velocity corrections. Goodness of fit uses $\chi^2_{\text{SN}} = \Delta\mu^{T} \mathcal{C}^{-1} \Delta\mu$, with $\mathcal{C}$ the $1048 \times 1048$ covariance matrix (statistical and systematic) of \cite{Scolnic_2018}. In each bin $H_0$ is inferred via MCMC with $\Omega_{m}^{0}$ fixed at the \cite{Scolnic_2018} value, varied within $2\sigma$ of the prior \cite{Dainotti2021apj-powerlaw}. Results are insensitive to the initial $H_0$: adopting $M = -19.245$ ($H_0 = 73.5$ km s$^{-1}$ Mpc$^{-1}$) \cite{Dainotti2021apj-powerlaw}, we fix $H_0$ in the first bin to set $M$, then hold $M$ throughout. The decreasing $H_0(z)$ trend is independent of the bin number \cite{Dainottigalaxies10010024}.

Let us now analyze two distinct models by addressing from Eq.(\ref{nii111})  $H^2(x)=H_{0}^{2}h^2 (\bar{\Omega}_{m}^{0} e^{3x}+1-\bar{\Omega}_{m}^{0})$ with the fiducial quantity $\bar{\Omega}_{m}^{0}=0.298$ \cite{Dainotti2021apj-powerlaw} (the value of $H_0$ is correspondingly taken as $73.5$ km s$^{-1}$ Mpc$^{-1}$). The first is the case (dubbed PL) of a power-law scaling (in the redshift $z$) of the Hubble constant, widely studied in 
\cite{Dainotti2021apj-powerlaw,Dainottigalaxies10010024,2025JHEAp..4800405D}, 
i.e.
\begin{align}
h_{PL}(x) = e^{-\alpha x}\;,
\label{nii99pl}
\end{align} 
where $\alpha$ is a parameter 
having order of magnitude of 
$10^{-2}$ (see also \cite{2023A&A...674A..45J}). We consider the best fit obtained in \cite{Dainotti2021apj-powerlaw}: $\alpha=0.016\pm0.009 $.
For the second model (dubbed OP), after testing a large family of regular asymptotic functions, we instead assume
\begin{align}
h_{OP}(x) = 1+ \beta (1-e^{-x})\;,
\label{niiiop}
\end{align}
Using the 40-bin reconstruction, we perform a nonlinear fit obtaining the best-fit value
\begin{align}\label{bestfit}
\beta=-0.0158 \pm  0.0059\;.
\end{align}
\begin{figure}[ht!]
\includegraphics[width=0.55\textwidth]{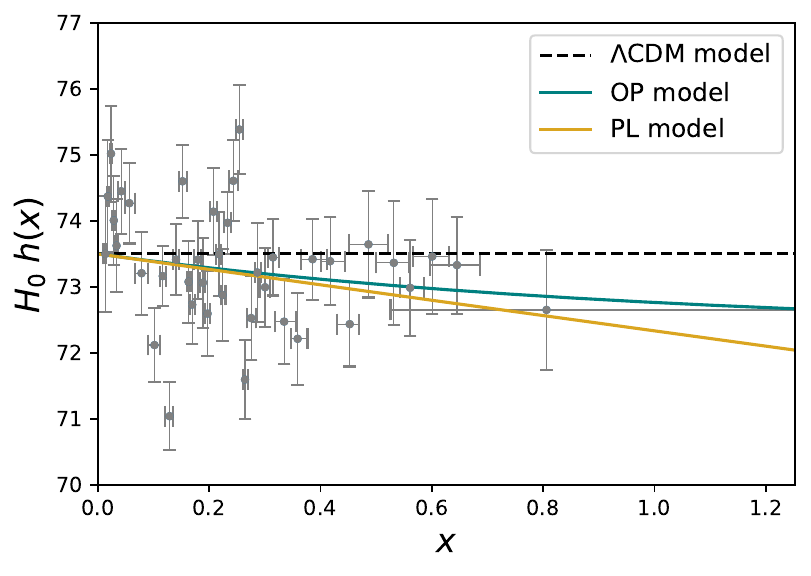}
\caption{Plot of $H_0 h(x)$ from Eq.(\ref{nii111}) and the fiducial values $H_0=73.5$ km s$^{-1}$ Mpc$^{-1}$, $\bar{\Omega}_{m}^{0}=0.298$: the yellow line represents the profile in Eq.(\ref{nii99pl}) with $\alpha=0.016$ (PL model), while the blue line correspond to Eq.(\ref{niiiop}) (OP model) with the best fit $\beta=-0.0158$. Bullets are $H_0$ data from \cite{Dainotti2021apj-powerlaw} with the corresponding error bars in 1$\sigma$ and bin widths in the $x$-axis. We also depict the constant line $h(x)=1$ for the base $\Lambda$CDM model (black dashed).}
\label{fig-H0}
\end{figure}

Both models are depicted in Fig.\ref{fig-H0}. Regarding the statistical performance, we obtain $\chi_{\text{red}}^2=2.047\, \text{(OP)},\,2.066\,\text{(PL)},\,2.176\,(\Lambda\text{CDM})$.
The large number of data points at low redshift, combined with the high number of degrees of freedom, leads to (inverse) p-values close to unity for all models. The ordering of the $\chi_{\text{red}}^2$ values indicates that the two models are almost statistically indistinguishable while they are both superior to the $\Lambda$CDM picture in reproducing the binned data (for physically motivated analysis of the PL model, see \cite{schiavone_mnras,Dainotti_2020,2021ApJ...914L..40D,Bargiacchi2023MNRAS.521.3909B,Dainotti2022PASJ...74.1095D,Dainotti2023ApJ...951...63D,Dainotti2023ApJ...950...45D,Lenart2023}). We also report the Akaike and Bayesian information criteria: AIC$=102.38\,\text{(OP)},\,103.12\,\text{(PL)},\,104.11\, (\Lambda\text{CDM})$ and
BIC$=104.07\,\text{(OP)},\,104.8\,\text{(PL)},\,104.11\, (\Lambda\text{CDM})$. 

\begin{figure}[ht!]
\includegraphics[width=0.55\textwidth]{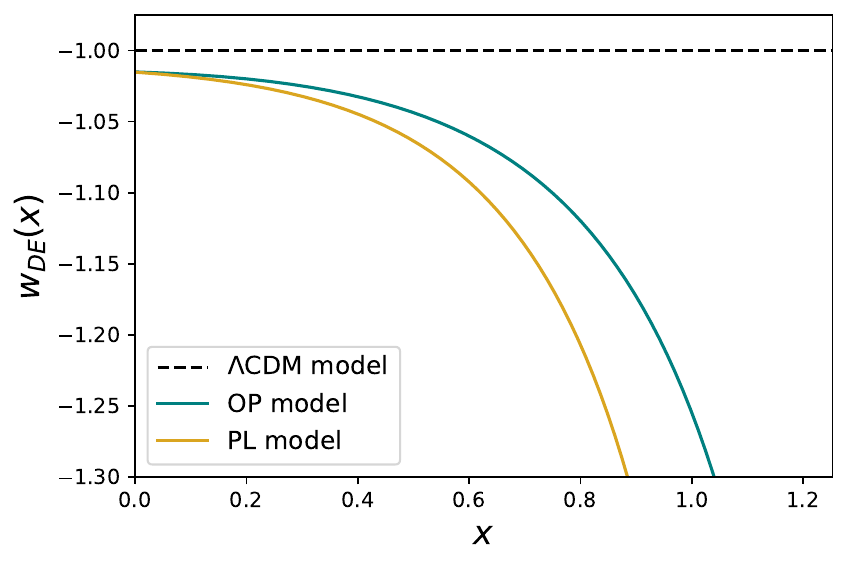}
\caption{Plot of $w_{DE}(x)$ from Eq.(\ref{nii37}): the PL model running function in Eq.(\ref{nii99pl}) is in yellow, while the OP scheme in Eq.(\ref{niiiop}) in blue. Black dashed represents the constant $\Lambda$CDM model $h(x)=1$ with $w_{DE}=-1$.}
\label{fig-wde}
\end{figure}

For the considered models, we can now apply the mapping and define the dynamical features of the dark energy component by using Eq.(\ref{nii37}) (together with Eq.(\ref{nii9876})). The resulting equation of state parameters are plotted in Fig.\ref{fig-wde}. The most interesting outcome is that the tested models exhibit a phantom behavior, with the dark energy equation of state below the cosmological constant threshold ($w_{\text{DE}} < -1$) over the probed range.

Our study has constructed a simple algorithm, able to link the 
functional form scaling a $\Lambda$CDM model with the equation of state of 
an evolutionary dark energy. The value of such an algorithm relies on the capability to account for the simultaneous observation of both the scaled dynamics (like the power-law behavior studied in \cite{2025JHEAp..4800405D}) and the evolutionary dark energy (emerging from the DESI Collaboration observation \cite{DESI:2025,DESI:2025zgx,2024ApJ...973L..14A}, when a CPL cosmological model 
\cite{Chevallier2001,Linder2003} is inferred; for related dynamical dark energy frameworks see also \cite{2000PhRvL..85.4438A,DiValentino:2022eot}). It is just the joint picture of these 
two different phenomenologies that constitutes the goal achieved by the present analysis, which offers a valuable interpretative framework for future observation of a possibly effective running Hubble constant.

\subsection*{Acknowledgment} M.G.D. acknowledges the support of the JSPS Grant-in-Aid for Scientific Research (KAKENHI) (A), Grant Number JP25H00675.


\begin{thebibliography}{10}
\ProvideTextCommand{\guillemotleft}{OT1}{%
  \leavevmode\raise .27ex\hbox{$\scriptscriptstyle\ll$}}
\ProvideTextCommand{\guillemotright}{OT1}{%
  \leavevmode\raise .27ex\hbox{$\scriptscriptstyle\gg$}}
\newcommand{\enquote}[1]{\guillemotleft#1\guillemotright}

\bibitem{2021CQGra..38o3001D}
Eleonora {Di Valentino}, Olga {Mena}, Supriya {Pan}, Luca {Visinelli}, Weiqiang {Yang}, Alessandro {Melchiorri}, David~F. {Mota}, Adam~G. {Riess} and Joseph {Silk}, \emph{Class. Quant. Grav.} \textbf{38}, 153001 (2021).

\bibitem{1998AJ....116.1009R}
Adam~G. {Riess}, Alexei~V. {Filippenko}, Peter {Challis}, Alejandro {Clocchiatti}, Alan {Diercks}, Peter~M. {Garnavich}, Ron~L. {Gilliland}, Craig~J. {Hogan}, Saurabh {Jha}, Robert~P. {Kirshner}, B.~{Leibundgut}, M.~M. {Phillips}, David {Reiss}, Brian~P. {Schmidt}, Robert~A. {Schommer}, R.~Chris {Smith}, J.~{Spyromilio}, Christopher {Stubbs}, Nicholas~B. {Suntzeff} and John {Tonry}, \emph{Astro. J.} \textbf{116}, 1009 (1998).

\bibitem{1999ApJ...517..565P}
S.~{Perlmutter}, G.~{Aldering}, G.~{Goldhaber}, R.~A. {Knop}, P.~{Nugent}, P.~G. {Castro}, S.~{Deustua}, S.~{Fabbro}, A.~{Goobar}, D.~E. {Groom}, I.~M. {Hook}, A.~G. {Kim}, M.~Y. {Kim}, J.~C. {Lee}, N.~J. {Nunes}, R.~{Pain}, C.~R. {Pennypacker}, R.~{Quimby}, C.~{Lidman}, R.~S. {Ellis}, M.~{Irwin}, R.~G. {McMahon}, P.~{Ruiz-Lapuente}, N.~{Walton}, B.~{Schaefer}, B.~J. {Boyle}, A.~V. {Filippenko}, T.~{Matheson}, A.~S. {Fruchter}, N.~{Panagia}, H.~J.~M. {Newberg}, W.~J. {Couch} and The Supernova~Cosmology {Project}, \emph{ApJ} \textbf{517}, 565 (1999).

\bibitem{Planck:2018vyg}
N.~Aghanim \emph{et~al.}, \emph{A\&A} \textbf{641}, A6 (2020).

\bibitem{Scolnic_2018}
D.~M. Scolnic and {et al.}, \emph{ApJ} \textbf{859}, 101 (2018).

\bibitem{Scolnic_2022}
D.~M. Scolnic and {et al.}, \emph{ApJ} \textbf{938}, 113 (2022).

\bibitem{Brout:2022vxf}
D.~Brout and {et al.}, \emph{ApJ} \textbf{938}, 110 (2022).

\bibitem{Dainotti2021apj-powerlaw}
Maria~Giovanna Dainotti, Biagio De~Simone, Tiziano Schiavone, Giovanni Montani, Enrico Rinaldi and Gaetano Lambiase, \emph{ApJ} \textbf{912}, 150 (2021).

\bibitem{arx2206.11447}
E.~{{\'O} Colg{\'a}in}, M.~M. {Sheikh-Jabbari}, R.~{Solomon}, M.~G. {Dainotti} and D.~{Stojkovic}, \emph{Phys. Dark Univ.} \textbf{44}, 101464 (2024).

\bibitem{DESI:2025}
A.G. {Adame} and {et al.}, \emph{JCAP} \textbf{2025}, 021 (2025).

\bibitem{DESI:2025zgx}
{DESI Collaboration}, \emph{Phys. Rev. D} \textbf{112}, 083515 (2025).

\bibitem{riess18}
Adam~G. {Riess}, Steven~A. {Rodney}, Daniel~M. {Scolnic}, Daniel~L. {Shafer}, Louis-Gregory {Strolger}, Henry~C. {Ferguson}, Marc {Postman}, Or~{Graur}, Dan {Maoz}, Saurabh~W. {Jha}, Bahram {Mobasher}, Stefano {Casertano}, Brian {Hayden}, Alberto {Molino}, Jens {Hjorth}, Peter~M. {Garnavich}, David~O. {Jones}, Robert~P. {Kirshner}, Anton~M. {Koekemoer}, Norman~A. {Grogin}, Gabriel {Brammer}, Shoubaneh {Hemmati}, Mark {Dickinson}, Peter~M. {Challis}, Schuyler {Wolff}, Kelsey~I. {Clubb}, Alexei~V. {Filippenko}, Hooshang {Nayyeri}, Vivian {U}, David~C. {Koo}, Sandra~M. {Faber}, Dale {Kocevski}, Larry {Bradley} and Dan {Coe}, \emph{ApJ} \textbf{853}, 126 (2018).

\bibitem{2021ApJ...908L...6R}
Adam~G. {Riess}, Stefano {Casertano}, Wenlong {Yuan}, J.~Bradley {Bowers}, Lucas {Macri}, Joel~C. {Zinn} and Dan {Scolnic}, \emph{ApJ Lett.} \textbf{908}, L6 (2021).

\bibitem{dainotti-desi}
E.~{{\'O.} Colg{\'a}in}, M.~G. {Dainotti}, S.~{Capozziello}, S.~{Pourojaghi}, M.~M. {Sheikh-Jabbari} and D.~{Stojkovic}, \emph{JHEAP} \textbf{49}, 100428 (2026).

\bibitem{2021ApJ...914L..40D}
M.~G. {Dainotti}, V.~{Petrosian} and L.~{Bowden}, \emph{ApJ Lett.} \textbf{914}, L40 (2021).

\bibitem{2020PhRvD.102b3518V}
Sunny {Vagnozzi}, \emph{\prd} \textbf{102}, 023518 (2020).

\bibitem{schiavone_mnras}
Tiziano Schiavone, Giovanni Montani and Flavio Bombacigno, \emph{Mon. Not. RAS} \textbf{522}, L72 (2023).

\bibitem{Montani:2023xpd}
Giovanni Montani, Mariaveronica De~Angelis, Flavio Bombacigno and Nakia Carlevaro, \emph{Mon. Not. RAS} \textbf{527}, L156 (2023).

\bibitem{Montani_2025}
G.~Montani, N.~Carlevaro and M.G. Dainotti, \emph{Phys. Dark Univ.} \textbf{48}, 101847 (2025).

\bibitem{Dainottigalaxies10010024}
Maria~Giovanna {Dainotti}, Biagio~De {De Simone}, Tiziano {Schiavone}, Giovanni {Montani}, Enrico {Rinaldi}, Gaetano {Lambiase}, Malgorzata {Bogdan} and Sahil {Ugale}, \emph{Galaxies} \textbf{10}, 24 (2022).

\bibitem{2025JHEAp..4800405D}
M.~G. {Dainotti}, B.~{De Simone}, A.~{Garg}, K.~{Kohri}, A.~{Bashyal}, A.~{Aich}, A.~{Mondal}, S.~{Nagataki}, G.~{Montani}, T.~{Jareen}, V.~M. {Jabir}, S.~{Khanjani}, M.~{Bogdan}, N.~{Fraija}, A.~C.~C. {do E.~S. Pedreira}, R.~H. {Dejrah}, A.~{Singh}, M.~{Parakh}, R.~{Mandal}, K.~{Jarial}, G.~{Lambiase} and H.~{Sarkar}, \emph{JHEAP} \textbf{48}, 100405 (2025).

\bibitem{fazzari2025}
E.~{Fazzari}, M.~G. {Dainotti}, G.~{Montani} and A.~{Melchiorri}, \emph{JHEAP} \textbf{49}, 100459 (2026).

\bibitem{arx2203.10558}
E.~{{\'O} Colg{\'a}in}, M.~M. {Sheikh-Jabbari}, R.~{Solomon}, G.~{Bargiacchi}, S.~{Capozziello}, M.~G. {Dainotti} and D.~{Stojkovic}, \emph{Phys. Rev. D.} \textbf{106}, L041301 (2022).

\bibitem{Colgain2021DDE}
Eoin {{\'O} Colg{\'a}in}, M.~M. {Sheikh-Jabbari} and Lu~{Yin}, \emph{\prd} \textbf{104}, 023510 (2021).

\bibitem{Schiavone:2022shz}
Tiziano Schiavone, Giovanni Montani, Maria~Giovanna Dainotti, Biagio De~Simone, Enrico Rinaldi and Gaetano Lambiase, \enquote{{Running Hubble constant from the SNe Ia Pantheon sample?}}, in \emph{{17th Italian-Korean Symposium on Relativistic Astrophysics}} (2022).

\bibitem{2024MNRAS.530.5091X}
Bing {Xu}, Jiancheng {Xu}, Kaituo {Zhang}, Xiangyun {Fu} and Qihong {Huang}, \emph{Mon. Not. RAS} \textbf{530}, 5091 (2024).

\bibitem{Liu2025PhRvD112L3539}
Tonghua Liu, Shuo Cao and Jieci Wang, \emph{Phys. Rev. D} \textbf{112}, L3539 (2025).

\bibitem{montaniEntropy}
Giovanni {Montani}, Elisa {Fazzari}, Nakia {Carlevaro} and Maria~Giovanna {Dainotti}, \emph{Entropy} \textbf{27}, 895 (2025).

\bibitem{2023PDU....4201348P}
Vivian {Poulin}, Tristan~L. {Smith} and Tanvi {Karwal}, \emph{Phys. Dark Univ.} \textbf{42}, 101348 (2023).

\bibitem{2019PhRvL.122v1301P}
Vivian {Poulin}, Tristan~L. {Smith}, Tanvi {Karwal} and Marc {Kamionkowski}, \emph{Phys. Rev. Lett.} \textbf{122}, 221301 (2019).

\bibitem{2023ARNPS..73..153K}
Marc {Kamionkowski} and Adam~G. {Riess}, \emph{Annual Review of Nuclear and Particle Science} \textbf{73}, 153 (2023).

\bibitem{2023PhRvD.108d3513H}
Laura {Herold} and Elisa G.~M. {Ferreira}, \emph{Phys. Rev. D} \textbf{108}, 043513 (2023).

\bibitem{2023Univ....9..393V}
Sunny {Vagnozzi}, \emph{Universe} \textbf{9}, 393 (2023).

\bibitem{DiValentino:2019ffd}
Eleonora Di~Valentino, Alessandro Melchiorri, Olga Mena and Sunny Vagnozzi, \emph{Phys. Dark Univ.} \textbf{30}, 100666 (2020).

\bibitem{DiValentinodark}
Eleonora {Di Valentino}, Ankan {Mukherjee} and Anjan~A. {Sen}, \emph{Entropy} \textbf{23}, 404 (2021).

\bibitem{Alestas2020PhRvD.101l3516A}
G.~{Alestas}, L.~{Kazantzidis} and L.~{Perivolaropoulos}, \emph{\prd} \textbf{101}, 123516 (2020).

\bibitem{Wang_2016}
B~Wang, E~Abdalla, F~Atrio-Barandela and D~Pavón, \emph{Reports on Progress in Physics} \textbf{79}, 096901 (2016).

\bibitem{MONTANI2024101486}
Giovanni Montani, Nakia Carlevaro and Maria~Giovanna Dainotti, \emph{Phys. Dark Univ.} \textbf{44}, 101486 (2024).

\bibitem{2025PDU....4901965D}
Eleonora {Di Valentino}, Jackson~Levi {Said}, Adam {Riess} and {et al.}, \emph{Phys. Dark Univ.} \textbf{49}, 101965 (2025).

\bibitem{Yang:2018uae}
Weiqiang Yang, Ankan Mukherjee, Eleonora Di~Valentino and Supriya Pan, \emph{Phys. Rev. D} \textbf{98}, 123527 (2018).

\bibitem{2025arXiv251220193M}
Giovanni {Montani}, Luis~A. {Escamilla}, Nakia {Carlevaro} and Eleonora {Di Valentino}, \emph{Phys. Rev. D} \textbf{113}, 023507 (2026).

\bibitem{NOJIRI201159}
Shin’ichi Nojiri and Sergei~D. Odintsov, \emph{Phys. Rept.} \textbf{505}, 59 (2011).

\bibitem{2007PhRvD..76f4004H}
Wayne {Hu} and Ignacy {Sawicki}, \emph{Phys. Rev. D} \textbf{76}, 064004 (2007).

\bibitem{2026JHEAp..5300612V}
A.~{Valletta}, G.~{Montani}, M.~G. {Dainotti} and E.~{Fazzari}, \emph{JHEAP} \textbf{53}, 100612 (2026).

\bibitem{2010JCAP...02..008S}
Daniel {Stern}, Raul {Jimenez}, Licia {Verde}, Marc {Kamionkowski} and S.~Adam {Stanford}, \emph{JCAP} \textbf{2010}, 008 (2010).

\bibitem{2018JCAP...04..051G}
Adri{\`a} {G{\'o}mez-Valent} and Luca {Amendola}, \emph{JCAP} \textbf{2018}, 051 (2018).

\bibitem{2018PhRvD..97l3507W}
Deng {Wang}, \emph{Phys. Rev. D} \textbf{97}, 123507 (2018).

\bibitem{Dainotti2024PDU....4401428D}
M.~G. {Dainotti}, A.~{L}. {Lenart}, M.~Ghodsi {Yengejeh}, S.~{Chakraborty}, N.~{Fraija}, E.~{Di Valentino} and G.~{Montani}, \emph{Phys. Dark Univ.} \textbf{44}, 101428 (2024).

\bibitem{Dainotti2024Galax..12....4D}
Maria~Giovanna {Dainotti}, Giada {Bargiacchi}, Aleksander~{\L}ukasz {Lenart} and Salvatore {Capozziello}, \emph{Galaxies} \textbf{12}, 4 (2024).

\bibitem{DAINOTTI202430}
M.G. Dainotti, G.~Bargiacchi, M.~Bogdan, S.~Capozziello and S.~Nagataki, \emph{JHEAP} \textbf{41}, 30 (2024).

\bibitem{2010A&A...523A...7G}
J.~{Guy} and {et al.}, \emph{A\&A} \textbf{523}, A7 (2010).

\bibitem{2011A&A...529L...4C}
N.~{Chotard} and {et al.}, \emph{A\&A} \textbf{529}, L4 (2011).

\bibitem{2023A&A...674A..45J}
X.~D. {Jia}, J.~P. {Hu} and F.~Y. {Wang}, \emph{A\&A} \textbf{674}, A45 (2023).

\bibitem{Dainotti_2020}
M.~G. Dainotti, A.~L. Lenart, G.~Sarracino, S.~Nagataki, S.~Capozziello and N.~Fraija, \emph{ApJ} \textbf{904}, 97 (2020).

\bibitem{Bargiacchi2023MNRAS.521.3909B}
G.~{Bargiacchi}, M.~G. {Dainotti}, S.~{Nagataki} and S.~{Capozziello}, \emph{Mon. Not. RAS} \textbf{521}, 3909 (2023).

\bibitem{Dainotti2022PASJ...74.1095D}
Maria~Giovanna {Dainotti}, Giuseppe {Sarracino} and Salvatore {Capozziello}, \emph{Publications of the Astronomical Society of Japan} \textbf{74}, 1095 (2022).

\bibitem{Dainotti2023ApJ...951...63D}
Maria~Giovanna {Dainotti}, Giada {Bargiacchi}, Malgorzata {Bogdan}, Aleksander~Lukasz {Lenart}, Kazunari {Iwasaki}, Salvatore {Capozziello}, Bing {Zhang} and Nissim {Fraija}, \emph{ApJ} \textbf{951}, 63 (2023).

\bibitem{Dainotti2023ApJ...950...45D}
M.~G. {Dainotti}, G.~{Bargiacchi}, A.~{L}. {Lenart}, S.~{Nagataki} and S.~{Capozziello}, \emph{ApJ} \textbf{950}, 45 (2023).

\bibitem{Lenart2023}
Aleksander~{\L}ukasz {Lenart}, Giada {Bargiacchi}, Maria~Giovanna {Dainotti}, Shigehiro {Nagataki} and Salvatore {Capozziello}, \emph{ApJ Supp. Series} \textbf{264}, 46 (2023).

\bibitem{2024ApJ...973L..14A}
{DESI Collaboration}, \emph{ApJ Letters} \textbf{973}, L14 (2024).

\bibitem{Chevallier2001}
M.~Chevallier and D.~Polarski, \emph{Int. J. Mod. Phys. D} \textbf{10}, 213 (2001).

\bibitem{Linder2003}
Eric~V. Linder, \emph{Phys. Rev. Lett.} \textbf{90}, 091301 (2003).

\bibitem{2000PhRvL..85.4438A}
C.~{Armendariz-Picon}, V.~{Mukhanov} and Paul~J. {Steinhardt}, \emph{Phys. Rev. Lett.} \textbf{85}, 4438 (2000).

\bibitem{DiValentino:2022eot}
Eleonora Di~Valentino, Nils~A. Nilsson and Mu-In Park, \emph{Mon. Not. RAS} \textbf{519}, 5043 (2023).

\end{thebibliography}

\end{document}